\documentclass[journal]{IEEEtran}
\usepackage{amsfonts}
\usepackage{amsmath}
\usepackage{amssymb}
\usepackage{subfigure}
\usepackage{epsfig}
\usepackage{color}
\usepackage{float}
\usepackage{cite,graphicx,algorithm}

\begin{document}

\title{Pilot Optimization and Power Allocation for OFDM-based Full-duplex Relay Networks with IQ-imbalances}
\author{{Jin Wang, Hai Yu, Yongpeng Wu, Feng Shu, Riqing Chen, Jun Li, and Jiangzhou Wang}
\thanks{This work was supported in part by the National Natural Science Foundation of China (Nos. 61472190, 61771244
 and 61702258), the Open Research Fund of National Key Laboratory of Electromagnetic Environment, China Research Institute of Radiowave Propagation (No. 201500013), the open research fund of National Mobile Communications Research Laboratory, Southeast University, China (No. 2013D02), the Research Fund for the Doctoral Program of Higher Education of China (No. 20113219120019), the Foundation of Cloud Computing and Big Data for Agriculture and Forestry (117-612014063).}
\thanks{Jin Wang, Hai Yu, Feng Shu, and Jun Li are with School of Electronic and Optical Engineering, Nanjing University of Science and Technology, 210094, China. E-mail:\{jin.wang, yuhai, shufeng\}@njust.edu.cn}
\thanks{Yongpeng Wu is with the Shanghai Key Laboratory of Navigation and Location-Based Services, Shanghai Jiao Tong University, Minhang 200240, China. E-mail: yongpeng.wu2016@gmail.com}
\thanks{Feng Shu, and Riqing Chen are with the College of Computer and Information Sciences, Fujian Agriculture and Forestry University, Fuzhou 350002, China. E-mail: riqing.chen@fafu.edu.cn}
\thanks{Jiangzhou Wang is with the School of Engineering and Digital Arts, University of Kent, Canterbury CT2 7NT, U.K. E-mail: j.z.wang@kent.ac.uk}
}
\maketitle
\begin{abstract}
In OFDM relay networks with IQ imbalances and full-duplex relay station (RS), how to optimize pilot pattern and power allocation using the criterion of minimizing the sum of mean square errors (Sum-MSE) for the frequency-domain least-squares channel estimator has a heavy impact on self-interference cancellation. Firstly, the design problem of pilot pattern is casted as a convex optimization. From the KKT conditions, the optimal analytical expression is derived given the fixed source power and RS power. Subsequently, an optimal power allocation (OPA) strategy is proposed and presented to further alleviate the effect of Sum-MSE under the total transmit power sum constraint of source node and RS. Simulation results show that the proposed OPA performs better than equal power allocation (EPA) in terms of Sum-MSE, and the Sum-MSE performance gain grows with deviating $\rho$ from the value of $\rho^o$ minimizing the Sum-MSE, where $\rho$ is defined as the average ratio of the residual SI channel at RS to the intended channel from source to RS. For example, the OPA achieves about 5dB SNR gain over EPA by shrinking or stretching $\rho$ with a factor $4$. More importantly, as $\rho$ decreases or increases more, the performance gain becomes more significant.
\end{abstract}


\section{Introduction}

With the help of full-duplex (FD) operation, cooperative relay networks can double the spectrum efficiency of the conventional relay network working in TDD/FDD way \cite{Sun,Sabharwal,Qi,Lix}. This is extremely important for the future wireless communications facing spectrum scarcity\cite{Masmoudi2,Xie,Lij}. The major challenge for a full-duplex transceiver is the strong self-interference (SI) from its own transmission \cite{Zhang,Koohian}. In \cite{Masmoudi}, the SI cancellation process is usually divided into two stages: radio-frequency (RF) cancellation and baseband cancellation. The RF cancellation is to significantly reduce the SI power, and the  baseband digital cancellation is to further remove the residual SI partially. For FD relay systems, how to provide a high-performance channel estimation by designing an appropriate channel estimator and optimizing pilot pattern and power allocation is crucial to  efficiently lower the effect of residual SI after RF cancellation \cite{Hu,Dongkyu}.

The authors in \cite{Xiong} proposed a maximum-likelihood (ML) channel estimator to simultaneously estimate both user-to-relay channels and SI channel at relay station (RS) in large-scale MIMO relay networks. To further achieve a reduction in the computational complexity of ML, the expectation-maximization (EM) iterative algorithm was adopted to implement ML. The proposed EM-based ML method showed a better performance than the arithmetic-mean-based one. In \cite{Li}, the Broyden-Fletcher-Goldfarb-Shanno algorithm was utilized to solve the ML estimator to estimate the intended and residual SI channel at destination in FD two-way relay systems, where pilot pattern is block-type. In practice, the existence of in-phase and quadrature (IQ) imbalance of OFDM transceivers makes it more complicated to estimate channels due to the destroyed orthogonality between subchannels \cite{Mokhtar,Liang,Tang}.

Channel estimation and pilot optimization in FD point-to-point OFDM systems with IQ imbalances were intensively investigated in \cite{Yuhai,Shufeng}. In \cite{Yuhai}, an adaptive orthogonal matching pursuit based channel estimator was proposed by exploiting the sparsity of both SI and intended channels mixed with IQ parameters. The proposed method performed much better than time-domain least-square (LS) due to exploiting the sparse property of channel. Two LS channel estimators were proposed and their optimal pilot patterns are formalised as a convex optimization problem in \cite{Shufeng}. When the transmit power of source is identical with that of destination, the close-form expression of optimal pilot product matrix was proved to be any four columns of an unitary matrix multiplied by a constant. In this paper, we extend this result to the FD relay networks with IQ imbalance. Here, RS operates in FD mode, and has unequal transmit power as source node. In such a more general scenario, power allocation becomes a challenging problem. Our main contributions are as follows:

$\bullet$ Fixing both transmit powers of source node and RS, in terms of minimum sum of mean square errors (Sum-MSE), where the two powers are equal or not equal,  pilot design is casted as a convex optimization. The optimal pilot pattern is derived for the frequency-domain LS channel estimator using the Karush-Kuhn-Tucker (KKT) conditions. When the transmit powers of source node and RS are identical, the optimal pilot pattern degenerates towards the special form in \cite{Shufeng}.

$\bullet$ Problem of power allocation is established as a geometric optimization. The optimal power allocation (OPA) strategy is derived and proposed under the total power sum constraint of source node and RS by using the Lagrangian multiplier method. Compared to equal power allocation, the proposed OPA shows a significant improvement in Sum-MSE performance as $\rho$ deviates far from its optimal feasible value of minimizing the Sum-MSE, where $\rho$ is defined as the average ratio of channel gain of the residual SI at RS to the that from source to RS and a positive number.

The remainder of this paper is organized as follows. Section II describes the full-duplex relay system model with IQ imbalance and the frequency-domain LS channel estimator is applied to for channel estimation. In Section III, the optimal pilot pattern and power allocation are derived to minimize the Sum-MSE. Simulation results are presented in Section IV. Finally, Section V concludes this paper.

\emph{Notations:} Matrices and vectors are denoted by letters of bold upper case and bold lower case, respectively. Signs $(\cdot)^H$, $(\cdot)^*$, $(\cdot)^T$, $(\cdot)^{-1}$ and $\text{tr}(\cdot)$ stand for the Hermitian conjugate, conjugate, transpose, inverse, and trace operation, respectively. The notation $\mathcal E\{\cdot\}$ and $\langle \cdot \rangle_N$ refer to the expectation and modulo operation. $\mathbf{I}_{N}$ denotes the $N \times N$ identity matrix and $\mathbf{0}_{N}$ denotes an all-zero matrix of size $N\times N$. $\otimes$ denotes the Kronecker product of two matrices. $\text{diag}\left\{\mathbf{a}\right\}$ represents a diagonal matrix formed by placing all elements of the vector $\mathbf{a}$ on its main diagonal.

\section{System model and channel estimator}
Fig.~1 sketches an OFDM-based decoding-and-amplifying (DF) relay network consisting of source node (S), destination node (D) and relay station (RS). It is assumed that there exists no direct link between source and destination. In this figure, the RS operates in FD mode. It receives the current frame of symbols transmitted from source, and at the same time sends the previous frame of symbols to destination over the same frequency band. $\mathbf{H}_{SR}\in \mathbb{C}^{N\times 1}$, $\mathbf{H}_{RD}\in \mathbb{C}^{N\times 1}$, and $\mathbf{H}_{RR}\in \mathbb{C}^{N\times 1}$ represent the intended source-to-relay, relay-to-destination, and the residual SI frequency-domain channel vectors from RS itself after RF cancellation, respectively, where $N$ denotes the total number of subcarriers. Assume all the channels are quasi-static Rayleigh fading, that is, all channel gain vectors keep constant during one frame. Here, one frame may include several hundreds or even thousands OFDM symbols. For the convenience of derivation and analysis below, block-type pilot pattern is adopted for channel estimation. Each frame includes $N_P$ successive pilot OFDM symbols and $N_D$ data OFDM symbols, where $N_P$ successive pilot OFDM symbols are placed in the beginning of each frame and $N_D\gg N_P$ such that a high spectrum efficiency is achieved.

\begin{figure}[ht]
  \centering
  \includegraphics[width=9cm]{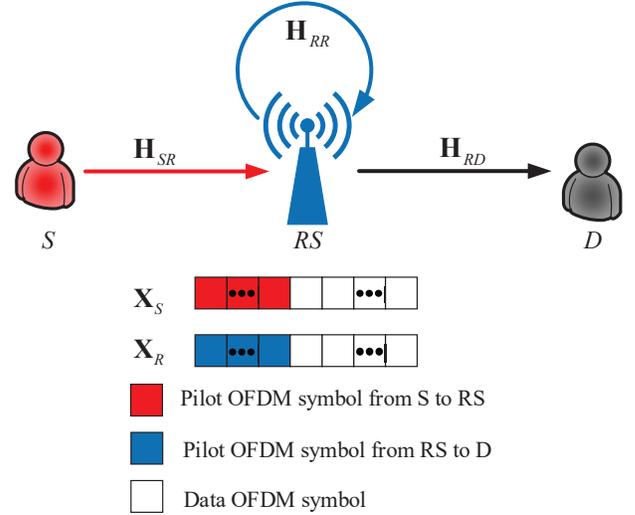}\\
  \caption{Block diagram of system model with block-type pilot pattern where S, RS, and D are short for source, relay station, and destination, respectively.}
\end{figure}

Since the imbalances between I and Q components at both the source transmitter and the RS transceiver generate the image of signals and destroy the orthogonality of subcarriers, the received symbol over subcarrier $k$ of OFDM symbol $n$ at relay station has the form
    \begin{align}
    \label{sys_mod_n_k}
y(n,k)&=H_{SR}^{a,n}(k)x_S(n,k)+H_{SR}^{b,n}(k)x_S^*(n,\hat{k})\nonumber\\
    &+\rho H_{RR}^{a,n}(k)x_R(n,k)+\rho H_{RR}^{b,n}(k)x_R^*(n,\hat{k})\nonumber\\
    &+\mu_{r,R}v(n,k)+\nu_{r,R}v^*(n,\hat{k}),
    \end{align}
where $\hat{k}=\langle N-k+2\rangle_N$ stands for the index of the image of subcarrier $k$. $x_S(n,k)$ and $x_D(n,k)$ denote the transmit pilot symbols from source node and RS corresponding to the $n$th OFDM symbol over $k$th subcarrier with $E\{|x_S(n,k)|^2\}=P_S$ and $E\{|x_R(n,k)|^2\}=P_R$, where $P_S$ and $P_R$ is the average signal power per subcarrier of source node and RS.
\begin{align}
H_{SR}^{a,n}(k)&=\mu_{r,R}\mu_{t,S}H_{SR}^n(k)+\nu_{r,R}\nu_{t,S}^*\left(H_{SR}^n(\hat{k})\right)^*,\\ H_{SR}^{b,n}(k)&=\mu_{r,R}\nu_{t,S}H_{SR}^n(k)+\nu_{r,R}\mu_{t,S}^*\left(H_{SR}^n(\hat{k})\right)^*,\\ H_{RR}^{a,n}(k)&=\mu_{r,R}\mu_{t,R}H_{RR}^n(k)+\nu_{r,R}\nu_{t,R}^*\left(H_{RR}^n(\hat{k})\right)^*,\\
H_{RR}^{b,n}(k)&=\mu_{r,R}\nu_{t,R}H_{RR}^n(k)+\nu_{r,R}\mu_{t,R}^*\left(H_{RR}^n(\hat{k})\right)^*,
\end{align}
and
\begin{align}\label{Source_IQ_mu}
\mu_{t,S}&=\cos(\theta_{t,S}/2)+j\alpha_{t,S}\sin(\theta_{t,S}/2),
\end{align}
\begin{align}\label{Source_IQ_nu}
\nu_{t,S}&=\alpha_{t,S}\cos(\theta_{t,S}/2)-j\sin(\theta_{t,S}/2),
\end{align}
where $\alpha_{t,S}$ and $\theta_{t,S}$ are amplitude and phase imbalances at transmitter of source node. Similar to (\ref{Source_IQ_mu}) and (\ref{Source_IQ_nu}), the  associated transmit and receive IQ-imbalance parameters at RS are defined as  $\mu_{t,R}$, $\nu_{t,R}$, $\mu_{r,R}$ and $\nu_{r,R}$, respectively. $v(n,k)$ is additive Gaussian white noise with zero mean and $\sigma_v^2$ variance in frequency domain at relay station.

It is particularly noted that the scalar parameter $\rho$ in  (\ref{sys_mod_n_k}) represents the average ratio of the residual SI channel gain to the intended channel gain, and reflects the relationship of who is dominant between the two channel gains. If $\rho>1$, then the residual SI channel is dominant and stronger than the intended channel. In other words, the useful messages are drowned in the residual self interference. If $\rho < 1$, then there is a converse result. That is, the intended signal is dominant over the residual SI channel. The value of $\rho$ depends on the relationship of RF SI cancellation capacity at RS and path loss from source to RS.  

The received symbol corresponding to subcarrier $\hat{k}$ is
\begin{align}
    \label{sys_mod_n_k1}
    y(n,\hat{k})&=H_{SR}^{a,n}(\hat{k})x_S(n,\hat{k})+H_{SR}^{b,n}(\hat{k})x_S^*(n,k)\nonumber\\
    &+\rho H_{RR}^{a,n}(\hat{k})x_R(n,\hat{k})+\rho H_{RR}^{b,n}(\hat{k})x_R^*(n,k)\nonumber\\
    &+\mu_{r,R}v(n,\hat{k})+\nu_{r,R}v^*(n,k).
\end{align}

In order to facilitate the following analysis, let us define $\mathbf{\Gamma}_{n,k}=[H_{SR}^{a,n}(k)~(H_{SR}^{b,n}(\hat{k}))^*~ H_{SR}^{b,n}(k)~(H_{SR}^{a,n}(\hat{k}))^*~H_{RR}^{a,n}(k)$ $(H_{RR}^{b,n}(\hat{k}))^*~H_{RR}^{b,n}(k)~(H_{RR}^{a,n}(\hat{k}))^*]^T$, $\mathbf{x}_{n,k}=[x_S(n,k)~x_S^*(n,\hat{k})~x_R(n,k)~x_R^*(n,\hat{k})]$, $\mathbf{A}=\text{diag}\{1,1,1,1,\rho,\rho,\rho,\rho\}$, $\mathbf{y}_{n,k}=[y(n,k)~y^*(n,\hat{k})]^T$, and $\mathbf{w}_{n,k}=[\mu_{r,R}v(n,k)+\nu_{r,R}v^*(n,\hat{k})~\mu^*_{r,R}v^*(n,\hat{k})+\nu^*_{r,R}v(n,k)]^T$. Stacking a pair of receive symbols over subcarriers $k$ and $\hat{k}$ forms the receive vector
\begin{align}
    \label{sys_mod_n_k2}
    \mathbf{y}_{n,k}=\left(\mathbf{x}_{n,k}\otimes\mathbf{I}_{2}\right)\mathbf{A}\mathbf{\Gamma}_{n,k}+\mathbf{w}_{n,k}.
\end{align}

In equation (\ref{sys_mod_n_k2}), there are eight unknowns but only two measurements. Eq. (\ref{sys_mod_n_k2}) is under-determined. Hence, at least $N_P\geq 4$ consecutive OFDM symbols are required to estimate $\mathbf{\Gamma}_{n,k}$ from (\ref{sys_mod_n_k2}). Stacking all the receive signals over subcarrier $k$ and $\hat{k}$ corresponding to these $N_P$ OFDM symbols yields
\begin{align}
    \mathbf{y}_k^P=\left(\mathbf{X}_k^P\otimes\mathbf{I}_{2}\right)\mathbf{A}\mathbf{\Gamma}_k+\mathbf{w}_k^P,
\end{align}
where $\mathbf{X}_k^P=[\mathbf{x}^T_{1,k}~\mathbf{x}^T_{2,k}~\cdots~\mathbf{x}^T_{N_P,k}]^T$,  $\mathbf{y}_k^P=[\mathbf{y}^T_{1,k}~\mathbf{y}^T_{2,k}~\cdots~\mathbf{y}^T_{N_P,k}]^T$, and $\mathbf{\Gamma}_k$ is  constant from pilot OFDM symbol $1$ to $N_P$ , thus  its subscript is omitted for convenience.  $\mathbf{w}_k^P=[\mathbf{w}^T_{1,k}~\mathbf{w}^T_{2,k}~\cdots~\mathbf{w}^T_{N_P,k}]^T$ with the covariance matrix being  $\mathbb{C}_w=E\{\mathbf{w}_k^P(\mathbf{w}_k^P)^H\}=\mathbf{I}_{N_P}\otimes \mathbf{C}_w$ and
\begin{align}
\mathbf{C}_{w}=\sigma^2_{v}
\left(\begin{array}{cc}
    |\mu_{r,R}|^2+|\nu_{r,R}|^2  & 2\mu_{r,R}\nu_{r,R}  \\
    2\mu_{r,R}^*\nu_{r,R}^*  & |\mu_{r,R}|^2+|\nu_{r,R}|^2
    \end{array}\right).
\end{align}

Given matrix $(\mathbf{X}_k^P)^H\mathbf{X}_k^P$ is invertible, the LS channel estimator is expressed as follows
\begin{align}
    \hat{\mathbf{\Gamma}}_k=\mathbf{A}^{-1}\left[\left((\mathbf{X}_k^P)^H\mathbf{X}_k^P\right)^{-1}(\mathbf{X}_k^P)^H\otimes\mathbf{I}_{2}\right]\mathbf{y}_k^P,
\end{align}
which gives the channel estimation error
\begin{align}\label{est_err}
    \Delta\hat{\mathbf{\Gamma}}_k=\mathbf{\Gamma}_k-\hat{\mathbf{\Gamma}}_k=\mathbf{A}^{-1}\left[\left((\mathbf{X}_k^P)^H\mathbf{X}_k^P\right)^{-1}(\mathbf{X}_k^P)^H\otimes\mathbf{I}_{2}\right]\mathbf{w}_k^P.
\end{align}

From (\ref{est_err}), we define the Sum-MSE corresponding to pilot subcarrier pair $k$
\begin{align}
   \text{Sum-MSE}_k&=E\{tr[\Delta\hat{\mathbf{\Gamma}}_k(\Delta\hat{\mathbf{\Gamma}}_k)^H]\}\\ \nonumber
   &=tr\{\mathbf{A}^{-2}[\left((\mathbf{X}_k^P)^H\mathbf{X}_k^P\right)^{-1}\otimes\mathbf{C}_w]\}.
\end{align}

Rewriting $\mathbf{A}=\mathbf{B}\otimes\mathbf{I}_{2}$ with $\mathbf{B}=\text{diag}\{1,1,\rho,\rho\}$ and using the property of Kronecker product computation\cite{Horn}, the above Sum-MSE is simplified as
\begin{align}\label{sum_mse_def}
   \text{Sum-MSE}_k=tr\{\mathbf{B}^{-2}\left((\mathbf{X}_k^P)^H\mathbf{X}_k^P\right)^{-1}\}tr\{\mathbf{C}_w\}.
\end{align}

\section{Optimal pilot design and power allocation}
In the previous section, an LS channel estimator and its Sum-MSE expression are  presented. In this section, by  minimizing its Sum-MSE, we attain its optimal pilot pattern in the convex optimization way. Then, the optimal power allocation policy  is casted as a geometric program subject to the total power sum constraint and computed by the KKT conditions.

\subsection{Optimal pilot pattern}
Firstly,  given  the transmit powers at source  and RS, the design problem of optimal pilot pattern is written as the following optimization
\begin{align}
\label{optim}
\min_{\mathbf{X}_k^P} ~~~&\text{Sum-MSE}_k\\ \nonumber
\text{s.t.} ~~~~&tr\{\mathbf{E}_S^H(\mathbf{X}_k^P)^H\mathbf{X}_k^P\mathbf{E}_S\}\leq 2N_PP_S,\\ \nonumber
&tr\{\mathbf{E}_R^H(\mathbf{X}_k^P)^H\mathbf{X}_k^P\mathbf{E}_R\}\leq 2N_PP_R,
\end{align}
with $\mathbf{E}_S=[\mathbf{I}_{2}~~\mathbf{0}_{2}]^H$ and $\mathbf{E}_R=[\mathbf{0}_{2}~~\mathbf{I}_{2}]^H$. Defining Gram matrix $\mathbf{Y}_k=(\mathbf{X}_k^P)^H\mathbf{X}_k^P$ and omitting the constant $tr\{\mathbf{C}_w\}$, the above optimization problem will be converted into
\begin{subequations}\label{optim2:1}
\begin{align}
\min_{\mathbf{Y}_k} ~~~&tr\{\mathbf{B}^{-2}\mathbf{Y}_k^{-1}\}\label{optim2:1a}\\
\text{s.t.} ~~~~&tr\{\mathbf{E}_S\mathbf{E}_S^H\mathbf{Y}_k\}\leq 2N_PP_S,\label{optim2:1b}\\
&tr\{\mathbf{E}_R\mathbf{E}_R^H\mathbf{Y}_k\}\leq 2N_PP_R,\label{optim2:1c}\\
&\mathbf{Y}_k\succ \mathbf{0}.\label{optim2:1d}
\end{align}
\end{subequations}

The Lagrangian dual function of (\ref{optim2:1}) is expressed as
\begin{align}
\mathcal{L}&(\mathbf{Y}_k,\lambda_k,\gamma_k,\mathbf{\Lambda}_k)\\ \nonumber
&=tr\{\mathbf{B}^{-2}\mathbf{Y}_k^{-1}\}+\lambda_k(tr\{\mathbf{E}_S\mathbf{E}_S^H\mathbf{Y}_k\}-2N_PP_S)
\\ \nonumber
&+\gamma_k(tr\{\mathbf{E}_R\mathbf{E}_R^H\mathbf{Y}_k\}-2N_PP_R)-tr\{\mathbf{\Lambda}_k\mathbf{Y}_k\},
\end{align}
where $\lambda_k\geq 0$, $\gamma_k\geq 0$, and $\mathbf{\Lambda}_k\succeq\mathbf{0}$ are the optimum dual variables associated with the constraints in (\ref{optim2:1b}), (\ref{optim2:1c}) and (\ref{optim2:1d})\cite{Boyd}. The KKT conditions related to $\mathbf{Y}_k$ are listed as
\begin{subequations}\label{kkt:1}
\begin{align}
&-\mathbf{Y}_k^{-1}\mathbf{B}^{-2}\mathbf{Y}_k^{-1}+\lambda_k\mathbf{E}_S\mathbf{E}_S^H+\gamma_k \mathbf{E}_R\mathbf{E}_R^H-\mathbf{\Lambda}_k=\mathbf{0},\label{kkt:1a}\\
&~~~~~~~~~~~~\lambda_k(tr\{\mathbf{E}_S\mathbf{E}_S^H\mathbf{Y}_k\}-2N_PP_S)=0,\label{kkt:1b}\\
&~~~~~~~~~~~~\gamma_k(tr\{\mathbf{E}_R\mathbf{E}_R^H\mathbf{Y}_k\}-2N_PP_R)=0,\label{kkt:1c}\\
&~~~~~~~~~~~~~~~~~~~~~~~~\mathbf{\Lambda}_k\mathbf{Y}_k=\mathbf{0}.\label{kkt:1d}
\end{align}
\end{subequations}

To guarantee $\mathbf{Y}_k\succ \mathbf{0}$, Eq.(\ref{kkt:1d}) holds only when $\mathbf{\Lambda}_k=\mathbf{0}$, and both $\lambda_k$ and $\gamma_k$ should be positive. Therefore,
\begin{align}
\mathbf{Y}_k\mathbf{B}^{2}\mathbf{Y}_k=\text{diag}\left\{\lambda_k^{-1},\lambda_k^{-1},\gamma_k^{-1},\gamma_k^{-1}\right\}.
\end{align}
%

Applying left and right multiplication by $\mathbf{B}$ to the above equation yields
\begin{align}
\left(\mathbf{B}\mathbf{Y}_k\mathbf{B}\right)^2=\text{diag}\left\{\lambda_k^{-1},\lambda_k^{-1},\rho^2\gamma_k^{-1},\rho^2\gamma_k^{-1}\right\}.
\end{align}


\emph{Lemma 1:}  For any diagonal matrix defined as $\mathbf{S}=\text{diag}\{s_1, s_2,\cdots,s_m,\cdots,s_M\}$ with $s_m>0$, there exist a unique Hermitian positive definite matrix $\mathbf{P}=\text{diag}\{\sqrt{s_1}, \sqrt{s_2}, \cdots, \sqrt{s_m},\cdots, \sqrt{s_M}\}$ satisfying $\mathbf{S}=\mathbf{P}^2$.

\emph{Proof:} See Appendix A.\hfill$\blacksquare$

Since $\mathbf{B}\mathbf{Y}_k\mathbf{B}$ is Hermitian positive definite, it has the following unique solution according to Lemma 1,
\begin{align}
\mathbf{B}\mathbf{Y}_k\mathbf{B}=\text{diag}\left\{\lambda_k^{-1/2},\lambda_k^{-1/2},\rho\gamma_k^{-1/2},\rho\gamma_k^{-1/2}\right\}.
\end{align}


Subsequently, we obtain
\begin{align}
\mathbf{Y}_k=\text{diag}\left\{\lambda_k^{-1/2},\lambda_k^{-1/2},\rho^{-1}\gamma_k^{-1/2},\rho^{-1}\gamma_k^{-1/2}\right\}.
\end{align}


Based on the complementary slackness condition, we obtain $tr\{\mathbf{E}_S\mathbf{E}_S^H\mathbf{Y}_k\}-2N_PP_S=0$ and $tr\{\mathbf{E}_R\mathbf{E}_R^H\mathbf{Y}_k\}-2N_PP_R=0$, it is derived that
\begin{align}
\lambda_k=\frac{1}{(N_PP_S)^2},\\
\gamma_k=\frac{1}{(\rho N_PP_R)^2}.
\end{align}

As a consequence,
\begin{align}\label{opt_y}
\mathbf{Y}_k=\text{diag}\left\{{N_PP_S},{N_PP_S},{N_PP_R},{N_PP_R}\right\}.
\end{align}

The above result can be summarized as the following theorem.

\emph{Theorem 1}: For an OFDM-based FD relay network in the presence of IQ imbalances, the optimal pilot matrix $\mathbf{X}_k^P$ should satisfy the optimality condition $(\mathbf{X}_k^P)^H\mathbf{X}_k^P=\text{diag}\left\{{N_PP_S},{N_PP_S},{N_PP_R},{N_PP_R}\right\}$ of minimizing the sum of MSE provided that the transmit powers $P_S$ and $P_R$ are fixed.
\hfill$\blacksquare$

\emph{Remark 1:} As $\mathbf{Y}_k=(\mathbf{X}_k^P)^H\mathbf{X}_k^P$, $\mathbf{X}_k^P$ can be constructed by any four orthogonal columns of an $N_P\times N_P$ unitary matrix multiplied by $\sqrt{N_PP_S}$, $\sqrt{N_PP_S}$, $\sqrt{N_PP_R}$ and $\sqrt{N_PP_R}$, respectively. Specially, for $k=1$ and $N/2+1$, the first column of $\mathbf{X}_k^P$ is conjugate to the second one, and the third column is conjugate to the forth one. Here, we use some special matrices to design their pilot symbols. Considering an $N_P\times N_P$ normalized discrete Fourier transform matrix, it is easy to find the column $m$ ($1<m\leq N_P$) and $N_P-m+2$ are conjugate and orthogonal with each other, from which the pilot matrix $\mathbf{X}_1^P$ and $\mathbf{X}_{N/2+1}^P$ can be well constructed. Taking $N_P=5$ for example, one of the optimal pilot matrix can be formed as
\begin{align}
\mathbf{X}_k^P=
\left(
                 \begin{array}{cccc}
                   \sqrt{P_S} & \sqrt{P_S} & \sqrt{P_R} & \sqrt{P_R} \\
                   \sqrt{P_S}W^1 & \sqrt{P_S}W^4 & \sqrt{P_R}W^2 & \sqrt{P_R}W^3\\
                   \sqrt{P_S}W^2 & \sqrt{P_S}W^8 & \sqrt{P_R}W^4 & \sqrt{P_R}W^6 \\
                   \sqrt{P_S}W^3 & \sqrt{P_S}W^{12} & \sqrt{P_R}W^8 & \sqrt{P_R}W^9 \\
                   \sqrt{P_S}W^4 & \sqrt{P_S}W^{16} & \sqrt{P_R}W^{10} & \sqrt{P_R}W^{12} \\
                 \end{array}
               \right)
\end{align}
with $W=e^{-j\frac{2\pi}{5}}$.

However, it doesn't make sense when $N_P=4$. Fortunately, we observe that, for a 4-order standard Hadamard matrix, multiplying its even rows by $j$, and each column by $\sqrt{P_S}$, $\sqrt{P_S}$, $\sqrt{P_R}$ and $\sqrt{P_R}$, a feasible form of pilot matrix will be shown as
\begin{align}
\mathbf{X}_k^P=\left(
                 \begin{array}{cccc}
                   \sqrt{P_S} & \sqrt{P_S} & \sqrt{P_R} & \sqrt{P_R} \\
                   j\sqrt{P_S} & -j\sqrt{P_S} & j\sqrt{P_R} & -j\sqrt{P_R} \\
                   \sqrt{P_S} & \sqrt{P_S} & -\sqrt{P_R} & -\sqrt{P_R} \\
                   j\sqrt{P_S} & -j\sqrt{P_S} & -j\sqrt{P_R} & j\sqrt{P_R} \\
                 \end{array}
               \right).
\end{align}
\hfill$\blacksquare$

Substituting (\ref{opt_y}) into (\ref{sum_mse_def}), we have the minimum Sum-MSE as follows
\begin{align}\label{sum_mse_min}
    \text{Sum-MSE}_k=\frac{2}{N_P}(\frac{1}{P_S}+\frac{1}{\rho^2P_R})tr(\mathbf{C}_w),
\end{align}

\subsection{Optimal power allocation}
Observing (\ref{sum_mse_min}), we find the minimum Sum-MSE relies heavily on the transmit power of the source and RS. Now, we turn to optimize the $P_S$ and $P_R$ under the condition $P_S+P_R\leq P$. This problem can be formulated as the following geometric program
\begin{align}
    \min_{P_S,P_R}~~~&\frac{2}{N_P}(\frac{1}{P_S}+\frac{1}{\rho^2P_R})tr(\mathbf{C}_w)\\ \nonumber
    \text{s.t.}~~~~~&P_S+P_R\leq P.
\end{align}

To solve the above convex optimization problem, we construct the associated Lagrangian function as
\begin{align}
\mathcal{L}(P_S,P_R)=\frac{2}{N_P}(\frac{1}{P_S}+\frac{1}{\rho^2P_R})tr(\mathbf{C}_w)+\xi(P_S+P_R-P)
\end{align}
where $\xi$ is the Lagrange multiplier. Setting the first-order derivative of the above function with respect to $P_S$ and $P_R$ to zero,
\begin{align}
\frac{\partial\mathcal{L}(P_S,P_R)}{\partial P_S}&=-\frac{2}{N_P}\frac{1}{P_S^2}tr(\mathbf{C}_w)+\xi=0,\\
\frac{\partial\mathcal{L}(P_S,P_R)}{\partial P_R}&=-\frac{2}{N_P}\frac{1}{\rho^2P_R^2}tr(\mathbf{C}_w)+\xi=0,
\end{align}
it is easy to obtain that $P_S=\rho P_R$ and $\xi\neq 0$. This yields $P_S+P_R-P=0$ in accordance with the complementary slackness condition, thus the optimal power allocation (OPA) of source and RS becomes
\begin{align}
{P_S}&=\frac{\rho P}{(1+\rho)},\\
{P_R}&=\frac{P}{(1+\rho)}.
\end{align}

This solution is concluded as the following theorem:

\emph{Theorem 2:} In OFDM-based FD relay networks with IQ imbalances, the optimal power allocation strategy of minimizing the Sum-MSE is given by ${P_S}=\frac{\rho P}{(1+\rho)}$ and ${P_R}=\frac{P}{(1+\rho)}$ subject to the total power constraint of source node and RS  $P_S+P_R\leq P$.
\hfill$\blacksquare$

Apparently, when $\rho>1$, more power is allocated to source node. And inversely, when $\rho<1$, RS takes up more power.

In this case, the corresponding minimum Sum-MSE of subcarrier $k$ and $\hat{k}$ becomes
\begin{align}
    \text{Sum-MSE}_k^o=\frac{2}{N_P}(1+\frac{1}{\rho})^2\frac{1}{P}tr(\mathbf{C}_w).
\end{align}
According to (3), the received SNR is defined as
\begin{align}
    \gamma=\frac{(|\mu_{t,S}|^2+|\nu_{t,S}|^2)P_S+\rho^2(|\mu_{t,R}|^2+|\nu_{t,R}|^2)P_R}{\sigma_v^2},
\end{align}
thus the minimum Sum-MSE can be expressed as
\begin{align}
    &\text{Sum-MSE}_k^o(\rho, \gamma)=\frac{4(|\mu_{r,R}|^2+|\nu_{r,R}|^2)}{\gamma N_P}\\ \nonumber
    &[(1+\frac{1}{\rho})(|\mu_{t,S}|^2+|\nu_{t,S}|^2)+(1+\rho)(|\mu_{t,R}|^2+|\nu_{t,R}|^2)].
\end{align}

The second derivative of the above minimum Sum-MSE with respect to $\rho$ is
\begin{align}
&\frac{\partial^2\text{Sum-MSE}_k^o(\rho, \gamma)}{\partial^2 \rho}\\ \nonumber
    &=\frac{8(|\mu_{r,R}|^2+|\nu_{r,R}|^2)(|\mu_{t,S}|^2+|\nu_{t,S}|^2)}{\rho^3 \gamma N_P}>0
\end{align}
for $\rho>0$, which means the minimum Sum-MSE is a convex function of $\rho$ for $\rho\in(0, +\infty]$ provided that SNR is fixed. In other words, the function $\text{Sum-MSE}_k^o(\rho, \gamma)$, with fixed variable SNR, has a globally minimum value in its domain. Setting the first-order derivative of minimum Sum-MSE with respect to $\rho$ to zero forms
\begin{align}
&\frac{\partial\text{Sum-MSE}_k^o(\rho, \gamma)}{\partial \rho}
=\frac{4(|\mu_{r,R}|^2+|\nu_{r,R}|^2)}{\gamma N_P}\\ \nonumber
    &~~~~[-\frac{1}{\rho^2}(|\mu_{t,S}|^2+|\nu_{t,S}|^2)+(|\mu_{t,R}|^2+|\nu_{t,R}|^2)]=0,
\end{align}
which yields
\begin{align}
    \rho^o=\sqrt{\frac{|\mu_{t,S}|^2+|\nu_{t,S}|^2}{|\mu_{t,R}|^2+|\nu_{t,R}|^2}}.
\end{align}
Finally, we obtain the globally minimum value of Sum-MSE
\begin{align}
    \text{Sum-MSE}_k^o&(\rho^o,\gamma)=\frac{4(|\mu_{r,R}|^2+|\nu_{r,R}|^2)}{\gamma N_P}\\ \nonumber &(\sqrt{|\mu_{t,S}|^2+|\nu_{t,S}|^2}+\sqrt{|\mu_{t,R}|^2+|\nu_{t,R}|^2}])^2
\end{align}

This result will be further verified in the next section.


\section{Simulation results}
In what follows, numerical simulation results are presented to evaluate the performance of proposed methods. The system parameters are set as follows: number of OFDM subcarriers $N=512$, length of cyclic prefix $L=32$, signal bandwidth $BW=10MHz$, number of pilot OFDM symbols $N_P=4$, carrier frequency $fc=2GHz$, and 16QAM is used for digital modulation.

In Figs.~2-4, the parameters of amplitude and phase imbalances between I and Q branches are chosen as $\alpha_{t,S}=5dB$, $\alpha_{t,R}=\alpha_{r,R}=1dB$, and $\theta_{t,S}=\theta_{t,R}=\theta_{r,R}=1^\circ$. For comparison, the equal power allocation (EPA) of source node and relay station is plotted as reference. The Sum-MSE corresponding to EPA is expressed as
\begin{align}
    &\text{Sum-MSE}^e_k(\rho, \gamma)=\frac{4(|\mu_{r,R}|^2+|\nu_{r,R}|^2)}{\gamma N_P}\\ \nonumber
    &[(1+\frac{1}{\rho^2})(|\mu_{t,S}|^2+|\nu_{t,S}|^2)+(1+\rho^2)(|\mu_{r,S}|^2+|\nu_{r,S}|^2)].
\end{align}

\begin{figure}[ht]
  \centering
  \includegraphics[width=8.8cm]{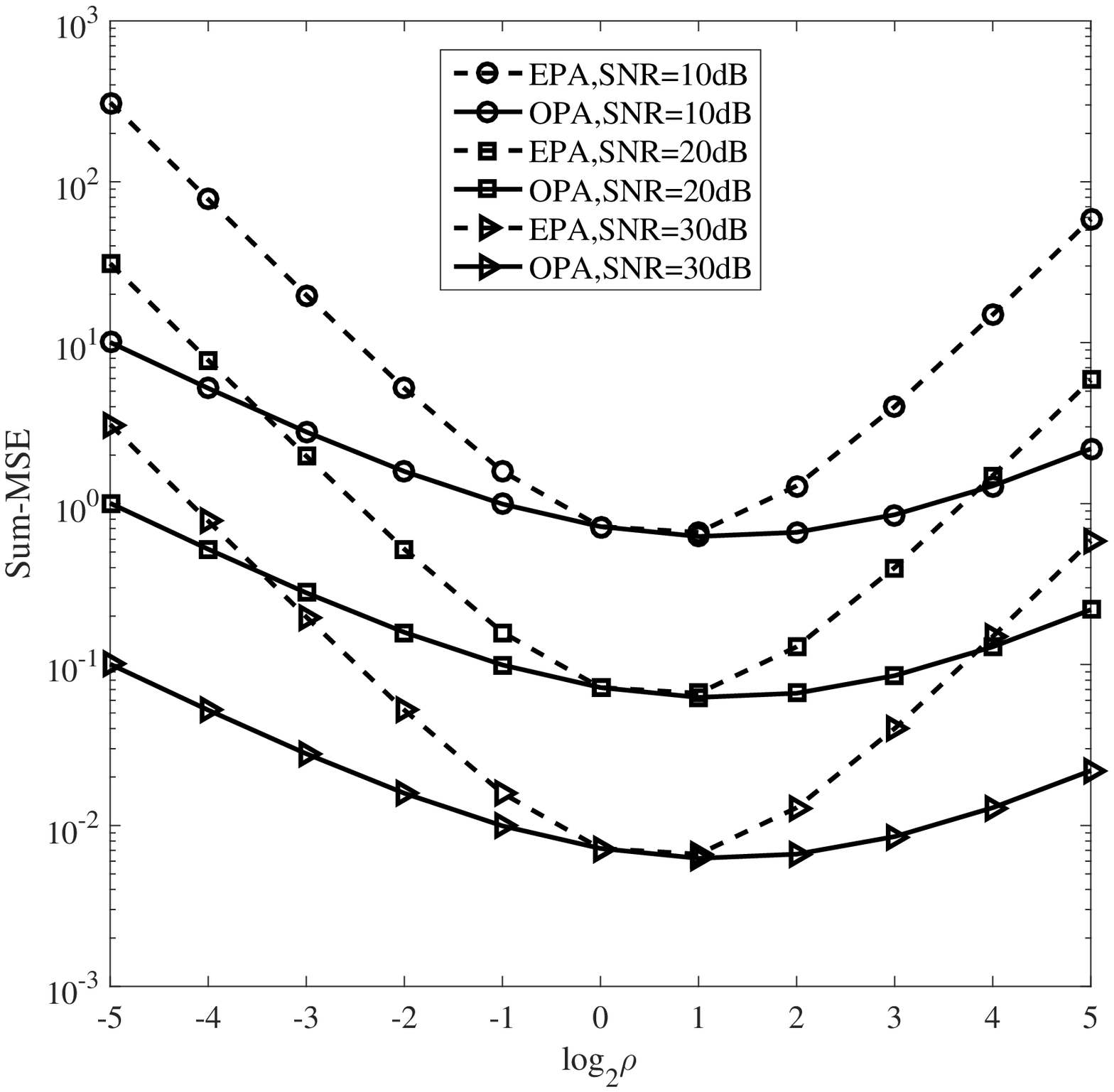}\\
  \caption{Sum-MSE versus $\rho$ for three typical receive SNRs ($\alpha_{t,S}=5dB$, $\alpha_{t,R}=\alpha_{r,R}=1dB$)}
\end{figure}
Fig.~2 demonstrates the curves of Sum-MSE versus $\rho$ of the proposed pilot pattern and power allocation for three typical receive SNRs. It is obvious that the proposed OPA performs better than EPA for all cases ($\rho>0$). Amazingly, as the value of $\rho$ is far away from $\rho^o$, the Sum-MSE gain achieved by OPA over EPA grows gradually. This implies that the larger performance benefit achieved by OPA is harvested by deviating the value of $\rho$ from $\rho^o$.

\begin{figure}[ht]
  \centering
  \includegraphics[width=8.8cm]{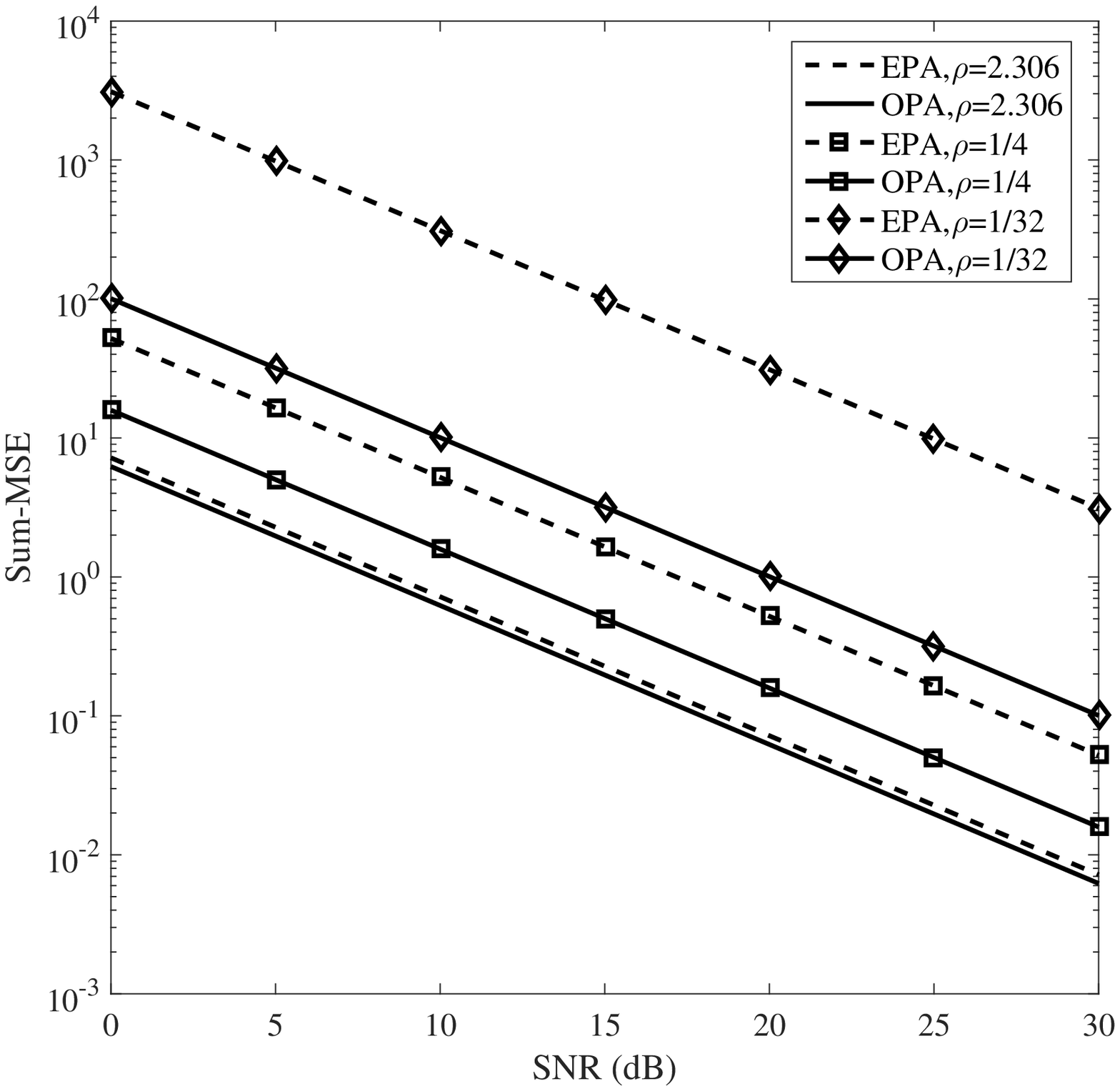}\\
  \caption{Sum-MSE versus receive SNR for three different values of $\rho$ ($\leq\rho^o$) ($\alpha_{t,S}=5dB$, $\alpha_{t,R}=\alpha_{r,R}=1dB$)}
\end{figure}

Fig.~3 displays the curves of Sum-MSE versus receive SNR of the proposed pilot pattern and power allocation for three different values of $\rho$ ($\leq\rho^o$). It is seen from this figure that a smaller $\rho$ leads to a larger Sum-MSE gain achieved by OPA over EPA. For example, OPA makes an approximate 5dB SNR gain over EPA when $\rho=1/4$, and 17dB when $\rho=1/32$. This trend can be explained by the fact that RS needs more power to improve the estimate accuracy of residual SI channel for a small $\rho$, i.e., a weak SI channel.

\begin{figure}[ht]
  \centering
  \includegraphics[width=8.8cm]{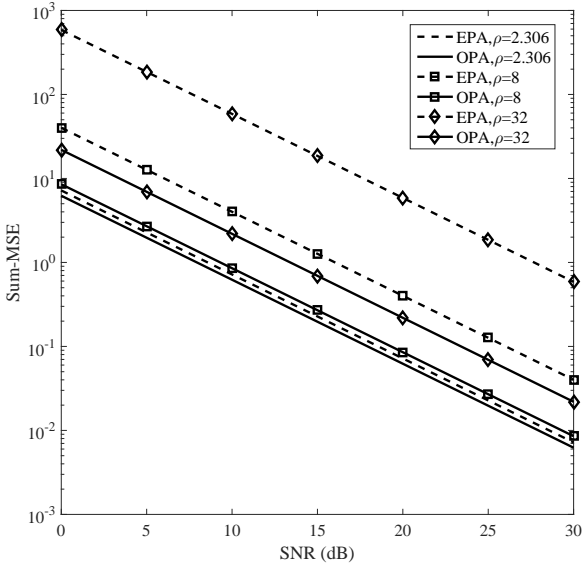}\\
  \caption{Sum-MSE versus receive SNR for for three different values of $\rho$ ($\geq\rho^o$) ($\alpha_{t,S}=5dB$, $\alpha_{t,R}=\alpha_{r,R}=1dB$)}
\end{figure}

Similar to Fig.~3, Fig.~4 shows the curves of Sum-MSE versus receive SNR of the proposed method for three different values of $\rho$ ($\geq\rho^o$). The Sum-MSE gain achieved by OPA over EPA becomes larger as $\rho$ increases. The OPA makes an about 7dB SNR gain over EPA when $\rho=8$, while it achieves 16dB SNR gain when $\rho=32$. The major reason is that a large $\rho$ means the SI channel is stronger than intended channel, hence, more power should be allocated to source node to enhance the estimate precision of intended channel.

In the following, from Fig.~5 to Fig.~7, we set the symmetric parameters of amplitude and phase imbalances between I and Q branches as $\alpha_{t,S}=\alpha_{t,R}=\alpha_{r,R}=1dB$, and $\theta_{t,S}=\theta_{t,R}=\theta_{r,R}=1^\circ$, thus $\rho^o=1$.

\begin{figure}[ht]
  \centering
  \includegraphics[width=8.8cm]{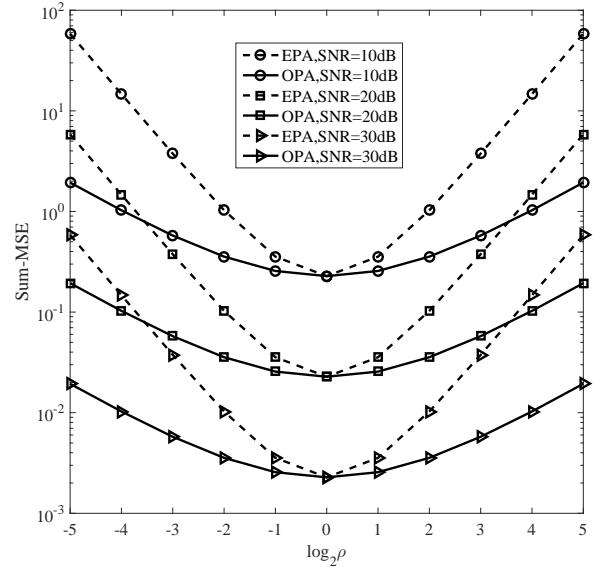}\\
  \caption{Sum-MSE versus $\rho$ for three typical receive SNRs ($\alpha_{t,S}=\alpha_{t,R}=\alpha_{r,R}=1dB$)}
\end{figure}

Fig.~5 plots the curves of Sum-MSE versus $\rho$ of the proposed method for three typical receive SNRs. As shown in Fig.~2, the proposed OPA performs better than EPA for all cases. And the Sum-MSE gain achieved by OPA over EPA grows gradually as the value of $\rho$ deviating from $1$. Due to the same parameters of IQ imbalances at source and RS transmitters, the curves of Sum-MSE versus $\rho$ are symmetric with respect to the line $\rho=1$.

\begin{figure}[ht]
  \centering
  \includegraphics[width=8.8cm]{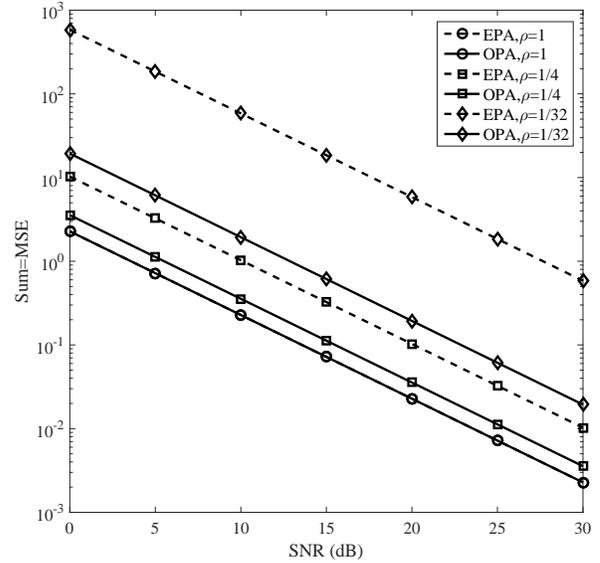}\\
  \caption{Sum-MSE versus receive SNR for three different values of $\rho$ ($\leq 1$) ($\alpha_{t,S}=\alpha_{t,R}=\alpha_{r,R}=1dB$)}
\end{figure}

Fig.~6 illustrates the curves of Sum-MSE versus receive SNR of the proposed pilot pattern and power allocation for three different values of $\rho$ ($\leq 1$). Both EPA and OPA achieve the same minimum Sum-MSE at $\rho=1$. Observing this figure, we find that the Sum-MSE gain achieved by OPA over EPA increases as $\rho$ decreases. Particularly, the proposed OPA attains an about 5dB SNR gain over EPA at $\rho=1/4$, and the SNR gain grows up to 15dB at $\rho=1/32$.

\begin{figure}[ht]
  \centering
  \includegraphics[width=8.8cm]{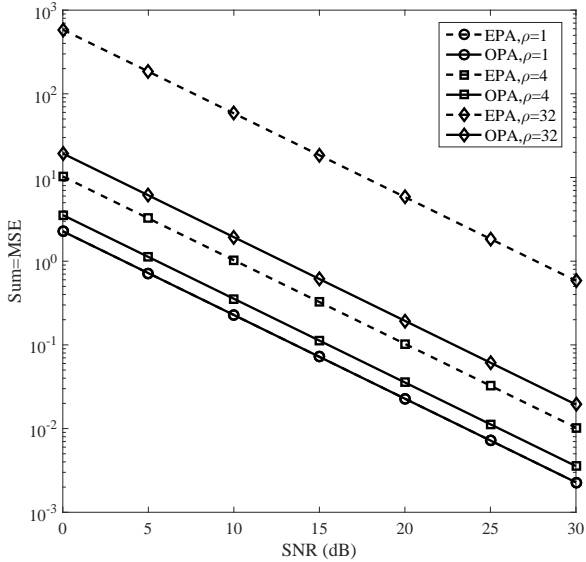}\\
  \caption{Sum-MSE versus receive SNR for three different values of $\rho$ ($\geq 1$) ($\alpha_{t,S}=\alpha_{t,R}=\alpha_{r,R}=1dB$)}
\end{figure}

Finally, Fig.~7 indicates the curves of Sum-MSE versus receive SNR of the proposed method for three different values of $\rho$ ($\geq 1$). From Fig.~7, it still follows that the Sum-MSE gain grows as $\rho$ increases. This figure further verifies the fact that the proposed OPA always performs better than EPA in terms of Sum-MSE performance.

In summary, different from EPA, in our OPA scheme, more power is allocated toward RS to enhance the estimate accuracy of the residual SI channel when $\rho<1$. Otherwise, more power is given to source node to enhance the estimate accuracy of residual SI channel when $\rho>1$.

\section{Conclusion}
In this paper, we make an investigation of pilot optimization and power allocation for the frequency-domain LS channel estimator in a full-duplex OFDM relay network with IQ imbalances. The analytical expression for optimum pilot product matrix is given by minimizing the Sum-MSE and utilizing the KKT conditions. Following this, the PA problem is formulated as a geometric optimization subject to the total power sum of source and RS. Finally, the optimal PA strategy is proposed and its closed-form solution is derived. Also, the Sum-MSE performance is proved to be a convex function of $\rho$, and has a minimum value. From simulation results, we find that the Sum-MSE performance of the proposed OPA is better than that of EPA.  With the value of $\rho$ deviating more from the minimum $\rho^o$, the Sum-MSE performance gain achieved by OPA over EPA increases gradually. In summary, the proposed PA can radically improve the Sum-MSE performance of the LS channel estimator compared to EPA in the case that $\rho$ approaches zero from right or tends to positive infinity.

\appendices
\section{Proof of Lemma 1}
\emph{Proof:} Let $\mathbf{Q}\neq\mathbf{P}$ be another Hermitian positive definite matrix satisfying $\mathbf{S}=\mathbf{Q}^2$. As $\mathbf{P}-\mathbf{Q}\neq \mathbf{0}$, there must exist a nonzero real eigenvalue $a$ and eigenvector $\mathbf{\xi}$ of $\mathbf{P}-\mathbf{Q}$ such that
\begin{align}
(\mathbf{P}-\mathbf{Q})\mathbf{\xi}=a\mathbf{\xi}.
\end{align}
Consequently,
\begin{align}
\mathbf{\xi}^H(\mathbf{P}^2-\mathbf{Q}^2)\mathbf{\xi}&=\mathbf{\xi}^H\mathbf{P}(\mathbf{P}-\mathbf{Q})\mathbf{\xi}+\mathbf{\xi}^H(\mathbf{P}-\mathbf{Q})\mathbf{Q}\mathbf{\xi}\\ \nonumber
&=a\mathbf{\xi}^H(\mathbf{P}+\mathbf{Q})\mathbf{\xi}=0.
\end{align}

Since $a\neq0$, the above equation only holds when $\mathbf{\xi}^H(\mathbf{P}+\mathbf{Q})\mathbf{\xi}=0$, which contradicts the assumption $\mathbf{P}$ and $\mathbf{Q}$ are positive definite. Thus, $\mathbf{Q}=\mathbf{P}$, which completes the proof of Lemma 1.\hfill$\blacksquare$

\bibliographystyle{IEEEtran}
\bibliography{IEEEabrv,FDIQ}

\end{document}